\documentclass[prd,twocolumn,english,preprintnumbers,amsmath,amssymb,nofootinbib,superscriptaddress,a4]{revtex4-1}

\usepackage[latin1]{inputenc}
\usepackage{graphicx}
\usepackage{color}

\usepackage{color}

\usepackage{bm}
\usepackage{bbm}
\usepackage{amsmath,amsfonts,latexsym}

\usepackage{dsfont}

\newcommand{\Tr}{\mathrm{Tr}}

\newcommand{\pat}{\partial_t}
\newcommand{\Eqref}[1]{Eq.~\eqref{#1}}

\newcommand{\dif}[1]{\ensuremath{\medspace \mbox{d} #1}}
\newcommand{\pdif}[2]{\ensuremath{\frac{\partial #1}{\partial #2}}}

\newcommand{\fdif}[2]{\ensuremath{\frac{\delta #1}{\delta #2}}}

\newcommand{\be}{\begin{equation}}
\newcommand{\ee}{\end{equation}}
\usepackage{babel}
\makeatother

\begin{document}

\title{Renormalization Flow of Axion Electrodynamics}
\author{Astrid Eichhorn$^{1,3}$, Holger Gies$^{1,2}$ and Dietrich Roscher}
\affiliation{Theoretisch-Physikalisches Institut, Friedrich-Schiller-Universit\"at Jena,  Max-Wien-Platz 1,
             D-07743 Jena, Germany, \\
             $^2$ Helmholtz Institut Jena,  Helmholtzweg 4, D-07743 Jena,
             Germany\\
             $^3$ Perimeter Institute for Theoretical Physics, 31 Caroline Street N, Waterloo, N2L 2Y5, Ontario, Canada
} 

\begin{abstract}
  We study the renormalization flow of axion electrodynamics, concentrating on
  the non-perturbative running of the axion-photon coupling and the mass of the axion (like) particle. Due to a non-renormalization property of the axion-photon vertex, the
  renormalization flow is controlled by photon and axion anomalous
  dimensions. {As a
  consequence,}  {momentum-independent} axion self-interactions are not induced by photon
  fluctuations. The {non-perturbative} flow towards the ultraviolet exhibits a
  Landau-pole-type  behavior, implying that the system has a scale of maximum
  UV extension and that the renormalized axion-photon coupling in the deep
  infrared is bounded from above. Even though gauge invariance guarantees that
  photon fluctuations do not decouple in the infrared, the
  renormalized {couplings} remain finite even in the deep
  infrared {and even for massless axions}. {Within our truncation, we also observe the existence of an exceptional RG trajectory, which is extendable to arbitrarily high scales, without being governed by a UV fixed point.}
\end{abstract}

\maketitle

\section{Introduction}

The existence of a fundamental pseudoscalar particle is strongly
motivated by the Peccei-Quinn solution to the strong CP problem
\cite{Peccei:1977hh}. The axion, being the pseudo-Nambu-Goldstone
boson of a {spontaneously as well as anomalously} broken axial U(1) symmetry \cite{Weinberg:1977ma}, receives
a small coupling to electromagnetism in generic models
\cite{Kim:1979if,Dine:1981rt}, very similar to the neutral pion in QCD. As a
consequence, the effective theory below the QCD scale contains photons
and axions as light fundamental degrees of freedom. Their interaction is
governed by a dimension five operator with a coupling being inversely
proportional to the ({presumably} high) scale of Peccei-Quinn symmetry
breaking. For reviews, see, e.g., \cite{Peccei:2006as}.

The resulting effective theory, axion electrodynamics, actually has a
wide range of applications. For instance, it also occurs in the
context of macroscopic media exhibiting the magneto-electric effect
\cite{Hehl:2007ut}, as well as in any hypothetical theory beyond the
standard model with axion-like degrees of freedom
\cite{Wilczek:1982rv,Chikashige:1980ui}, and, of course, for the
description of pion decay into two photons.

In the present work, we take a more general viewpoint on axion
electrodynamics and study its renormalization properties in the
effective framework with pure photon and axion degrees of
freedom. There are several motivations for this study: first,
experiments actively searching for the axion cover a wide range of
scales: solar observations such as CAST \cite{CAST} or the Tokio
helioscope \cite{Moriyama:1998kd} search for axions emitted by the sun
through the Primakoff process with a typical keV momentum scale
\cite{Dicus:1978fp}. Even higher momentum scales are probed by the
stellar cooling rate in the Helium burning phase of horizontal branch
stars { and other astrophysical considerations as reviewed in
  \cite{Raffelt:1996wa,Nakamura:2010zzi,Hewett:2012ns}}, {with hints
  of possible observable effects even at the TeV scale
  \cite{Horns:2012kw}}. By contrast, purely laboratory based
experiments as the classical light-shining-through-walls (LSW)
\cite{Sikivie:1983ip,Cameron:1993mr,Robilliard:2007bq,Chou:2007zzc,Pugnat:2006ba,Afanasev:2008jt,Ehret:2010mh}, 
see also \cite{Redondo:2010dp} for an overview, or polarimetry set-ups
\cite{Cameron:1993mr,Zavattini:2007ee,Chen:2006cd} using optical
lasers work with momentum transfers on the $\mu$eV scale. Searches
which are sensitive to axion (like) dark matter operate at similar
scales \cite{Asztalos:2009yp}.  A sizable running of the couplings
over these nine orders of magnitude would have severe implications for
the comparison of experimental results \cite{Jaeckel:2006xm}. Even
though the coupling is expected to be weak, photon and axion
fluctuations could potentially lead to sizable renormalization
effects, because of their small mass. In particular, photon
fluctuations strictly speaking never decouple as their masslessness is
granted by gauge invariance.

Another motivation is of more conceptual type: even though QCD-type
models of axion electrodynamics do have a physical cut-off,
{approximately given by the scale of chiral symmetry breaking $\sim
\mathcal O(1)$GeV}, it remains an interesting question as to whether
axion electrodynamics could be a self-consistent fundamental quantum field
theory. {Within a nonperturbative context, the negative mass dimension of the axion-photon coupling does not prohibit the existence of the theory as a fundamental theory, it only precludes the possibility of the coupling becoming asymptotically free.}
If {the theory can only exist as an effective theory} -- as naively expected -- its breakdown in the ultraviolet
can put restrictions on the physically admissible values of the coupling and
the axion mass, much in the same way as the Higgs boson mass in the standard
model is bounded from above by UV renormalization arguments, see,
e.g., \cite{Hambye:1996wb}. 

Since the axion-photon-interaction as a dimension five operator is
perturbatively non-renormalizable, a nonperturbative treatment is
required. This work is thus based on the functional renormalization group
(RG) allowing to explore a controlled approach to the potentially strongly coupled
UV.

The motivating questions are answered by the present study {with
  little surprise, though with interesting lessons to be learned from
  field theory}: in the physically relevant weak-coupling parameter
regime, the renormalization effects remain tiny despite the presence
of massless fluctuations. Furthermore, no indications for {generic} UV completeness are found
beyond perturbation theory. Nevertheless, the present study yields an
interesting example of renormalization theory revealing several
unusual properties: we identify a nonrenormalization property of the
axion self-interactions and show that the RG flow to lowest order is
determined solely by the anomalous dimensions of the photon and the
axion field. {In addition to explicit solutions to the RG
  equations, we also identify an exceptional RG trajectory which
  exemplifies a new class of potentially UV-controlled flows.}

Despite the non-decoupling of massless modes in the deep infrared, our
RG flow predicts that the physical couplings reach finite IR
values. {Explicit solutions demonstrate, how the massless modes
  effectively decouple by means of a powerlaw}. Finally, the
{generic} UV incompleteness of axion-electrodynamics puts an
upper bound on the axion-photon interaction for small axion masses
much in the same way as the Higgs boson mass is bounded from above in
the standard model.

Our article is organized as follows: {in Sect. \ref{sec:II}, we
introduce axion electrodynamics in Minkowski and Euclidean
space. Section \ref{sec:III} is devoted to the quantization of axion
electrodynamics using the Wetterich equation, i.e., the RG flow
equation for the 1PI generating functional. The RG flow is
investigated in the theory space spanned by the axion mass and
axion-photon coupling in Sect.~\ref{sec:IV}, where the general
structure of the flow is discussed, as well as explicit solutions are
worked out. Phenomenological implications in the context of the QCD
axion as well as for more general axion-like particles (ALPs) are
summarized in Sect.~\ref{sec:implications}. We conclude in
Sect.~\ref{sec:conclusions} and defer some technical details
(App.~\ref{app:A}) as well as further conceptual considerations
(App.~\ref{app:B}) to the appendices.}

\section{Axion electrodynamics}
\label{sec:II}

We consider an axion-photon field theory with classical Lagrangian in
Minkowski space (with metric $\eta=(+,-,-,-)$) 
\begin{equation}
\mathcal{L}= -\frac{1}{4} F_{\mu\nu} F^{\mu\nu} - \frac{1}{2}
\partial_\mu a \partial^\mu a -\frac{1}{2} \bar{m}^2 a^2 
- \frac{1}{4} \bar{g} a F_{\mu\nu}\widetilde{F}^{\mu\nu}, \label{eq:LAED}
\end{equation}
where $\bar{m}$ denotes the axion mass and $\bar{g}$ the axion-photon
coupling. The latter has an inverse mass dimension.  This Lagrangian
-- considered as the definition of an effective classical field theory
-- serves as a starting point for a variety of axion-photon phenomena
such as the Primakoff effect \cite{Dicus:1978fp} or axion-photon
oscillations giving rise {to light-shining-through-wall
  signatures \cite{Sikivie:1983ip}, polarimetry effects
  \cite{Maiani:1986md,Ahlers:2006iz} or higher-frequency generation
  \cite{Dobrich:2010hi}}. For the QCD axion, the mass and coupling
parameters are related, $\bar{g} \sim \bar{m}$. {Particular
  implications for the example of a QCD axion will be discussed in
  Sect.~\ref{sec:implications}. In the following, we take a different
  viewpoint and} consider this Lagrangian as a starting point for {an
  (effective)} quantum field theory.

Let us carefully perform the rotation to Euclidean space where the
fluctuation calculation will ultimately be performed. We use the
standard conventions
\begin{eqnarray}
x^0|_\text{M} &=&-i x_4|_\text{E}, \quad A^0|_\text{M} =-i
A_4|_\text{E}, \quad \partial^0 |_\text{M} = i \partial_4
|_\text{E},\nonumber\\
 \mathcal{L}|_\text{M}&=& -\mathcal{L}|_\text{E},\label{eq:conv}
\end{eqnarray}
where $|_\text{M/E}$ marks the quantities in Minkowski/Euclidean
space, respectively. As a consequence, we get, for instance,
$\partial_\mu a \partial^\mu a|_\text{M} = - \partial_\mu a
\partial_\mu a|_\text{E}$, and $F_{\mu\nu}F^{\mu\nu}|_\text{M} =
F_{\mu\nu} F_{\mu\nu}|_\text{E}$, as usual. The axion-photon coupling
contains the structure $aF_{\mu\nu} \widetilde{F}^{\mu\nu} =
\frac{1}{2} a
\epsilon_{\mu\nu\kappa\lambda}F^{\mu\nu}F^{\kappa\lambda}$, which
includes $\partial^0 A^1 \partial^2 A^3|_\text{M}$ and
index permutations thereof as building blocks. As only one time-like
component appears, the Minkowskian and Euclidean versions of the
axion-photon couplings differ by a factor of $i$. A possible further
but irrelevant factor of $(-1)$ may or may not arise depending on the
conventions for the Levi-Civita symbol. 

The resulting Euclidean Lagrangian finally reads
\begin{equation}
\mathcal{L}|_\text{E}= \frac{1}{4} F_{\mu\nu} F_{\mu\nu} + \frac{1}{2}
\partial_\mu a \partial_\mu a +\frac{1}{2} \bar{m}^2 a^2 
+ \frac{1}{4} i \bar{g} a F_{\mu\nu}\widetilde{F}_{\mu\nu}, \label{eq:LAEDE}
\end{equation}
For a computation of photon fluctuations in the continuum, gauge
fixing is necessary. As there are no minimally coupled charges in pure
axion electrodynamics, we can perform a complete Coulomb-Weyl
gauge-fixing procedure, imposing the gauge conditions
\begin{equation}
\nabla\cdot \mathbf{A}=0, \quad A_0=0, \label{eq:gf}
\end{equation}
such that only the two transversal degrees of freedom of the
photon field remain. The corresponding gauge-fixing action in Minkowski space reads
\begin{eqnarray}
\mathcal{L}_{GF} &=&
\frac{1}{2\alpha}(\eta^{\mu\nu}-n^{\mu}n^\nu)(\partial_\mu A_\nu)
(\partial^\kappa A^\lambda)(\eta_{\kappa\lambda}-n_\kappa
n_\lambda) \nonumber\\
&&+\frac{1}{2\beta}(n_\mu A^\mu)^2, \label{eq:Lgf} 
\end{eqnarray} 
where $n_\mu=(1,0,0,0)$. The translation to Euclidean space using
\Eqref{eq:conv} is straightforward. The two gauge-parameters
$\alpha,\beta$ can be chosen independently. In the present work, we
will exclusively consider a Landau-gauge limit, $\alpha,\beta\to 0$,
corresponding to an exact implementation of the gauge-fixing
conditions.

\section{Quantization of axion electrodynamics}
\label{sec:III}

As the axion-photon coupling has an inverse mass dimension, i.e.,
$[\bar{g}]=-1$, axion electrodynamics does not belong to the class of
perturbatively renormalizable theories. Without any further
prerequisites, quantum fluctuations in this theory can still be dealt
with consistently in the framework of effective field theories. This
may potentially require the fixing of further physical parameters
corresponding to higher-order operators. If the theory featured a
non-Gau\ss ian UV fixed point with suitable properties, axion
electrodynamics could even be asymptotically safe
\cite{Weinberg:1976xy,Niedermaier:2006wt} and thus non-perturbatively
renormalizable.

A suitable framework to deal with quantum fluctuations in either
scenario is the functional renormalization group. It allows us to
study the renormalization flow of rather general classes of theories
specified in terms of their degrees of freedom and their
symmetries. For this, a convenient tool is the Wetterich equation \cite{Wetterich:1993yh}
\begin{equation}
\partial_t\Gamma_k=\frac{1}{2}\Tr\, \left[\partial_t
  R_k\left(\Gamma_k^{(2)} + R_k\right)^{-1}\right], \,\,\partial_t= k \frac{d}{dk},
\label{eq:Wetteq}
\end{equation}
for the effective average action $\Gamma_k$, which interpolates
between a microscopic or bare UV action
$S_{\Lambda}=\Gamma_{k\to\Lambda}$ and the full quantum effective
action $\Gamma=\Gamma_{k\to0}$.  The effective average action
$\Gamma_k$ governs the dynamics of the field expectation values after
having integrated out quantum fluctuations from a UV scale $\Lambda$
down to the infrared scale $k$.  The {infrared} regulator function $R_k$
specifies the details of the regularization of quantum fluctuations to
be integrated out near an infrared momentum shell with momentum
$k$. $\Gamma^{(2)}_k$ denotes the second functional derivative of the
effective average action with respect to the fields, and the trace
contains a summation/integration over all discrete/continuous indices,
reducing to a momentum integral in the simplest case. Thus the
Wetterich equation is a one-loop equation from a technical point of
view, but nevertheless includes effects at higher loop order in
perturbation theory since it is the full non-perturbative propagator
that enters the loop diagram. Since its derivation does not rely on
the existence of a small parameter, it is applicable also in the
non-perturbative regime.  For reviews of the functional RG see, e.g.,
\cite{Berges:2000ew}.

In the present work, we study the renormalization flow of axion
electrodynamics as parameterized by a class of (Euclidean) action functionals of
the form
\begin{eqnarray}
\Gamma_k &=& \int d^4{x}\left[\frac{Z_F}{4}(F_{\mu\nu}(x))^2+
  \frac{Z_a}{2}(\partial_\mu
  a(x))^2+\frac{\bar{m}_k^2}{2}a(x)^2\right.\nonumber\\
&& \left. \qquad
  +\frac{i\bar{g}_k}{4}a(x)F_{\mu\nu}(x)\tilde{F}_{\mu\nu}(x)\right]+Z_{F}\int
d^4x \mathcal{L}_{GF}, \label{eq:trunc} 
\end{eqnarray}
where we allow the mass and the coupling constant to be scale
dependent. Also, the wave function renormalizations $Z_{F/a}$ are
implicitly assumed to be scale dependent.  In principle, also the
gauge-fixing parameters $\alpha,\beta$ could run with $k$. However, we
use the known fact that the Landau-gauge limit is a fixed point
of the RG \cite{Ellwanger:1995qf}. {Further terms beyond this simple truncation of full axion-electrodynamics are induced by the operators present in our truncation, even if set to zero at a UV scale $\Lambda$,
and are generically expected to couple back into the flow of the couplings considered here.}

The flow equation involves the second functional derivative of the
action $\Gamma_k^{(2)}$, corresponding to the inverse propagator,
which is matrix-valued in field space,
\begin{equation}
\label{GammaMatrix}
\Gamma_k^{(2)} = \begin{pmatrix} \Gamma^{aa} & \Gamma^{aA} \\ \Gamma^{Aa} & \Gamma^{AA}\end{pmatrix}\,.
\end{equation}
In momentum space, the corresponding components read 
\begin{widetext}
\begin{subequations}
\label{GammaKomp}
\begin{align}
\Gamma^{aa} &=\fdif{^2\Gamma_k}{a(p)\delta a(-q)}=(2\pi)^{4}\delta^{(4)}(q-p)[{\bar{m}}_k^2+Z_aq^2]\,, \label{Gammaaa}\\
\Gamma_\lambda^{aA} &= \fdif{^2\Gamma_k}{a(p)\delta A_\lambda(-q)} = i{\bar{g}}_k\varepsilon^{\mu\nu\gamma\delta}\delta_{\nu\lambda}A_{\delta}(q-p)q_\mu(q_\gamma-p_\gamma)\,, \label{GammaaA}\\
\Gamma_\kappa^{Aa} &= \fdif{^2\Gamma_k}{A_\kappa(p)\delta a(-q)} = i{\bar{g}}_k\varepsilon^{\mu\nu\gamma\delta}\delta_{\nu\kappa}A_{\delta}(q-p)p_\mu(p_\gamma-q_\gamma)\,,\label{GammaAa}\\
\Gamma_{\kappa\lambda}^{AA} &=\fdif{^2\Gamma_k}{A_\kappa(p)\delta
  A_\lambda(-q)}\nonumber\\
& = i{\bar{g}}_k\varepsilon^{\mu\nu\gamma\delta}\delta_{\nu\lambda}\delta_{\delta\kappa}q_\mu p_\gamma a(q-p)
+(2\pi)^4\delta^{(4)}(q-p)Z_F\left[(q^2\delta_{\kappa\lambda}-q_\kappa q_\lambda)+(\delta_{\kappa\mu}-n_\kappa n_\mu)\frac{q_\mu q_\nu}{\alpha}(\delta_{\nu\lambda}-n_\nu n_\lambda)+\frac{n_\kappa n_\lambda}{\beta}\right]. \label{GammaAA}
\end{align}
\end{subequations}
Note that $\Gamma^{aa}$ and the second term of $\Gamma^{AA}$ denote
the field-independent inverse propagators, whereas
$\Gamma^{aA}$, $\Gamma^{Aa}$ and the first term of $\Gamma^{AA}$
correspond to vertices, since they contain powers of the external fields. 
The regulator $R_k$ is chosen diagonal in field space with the
components,
\begin{equation}
R_k^{aa}(q) = Z_a q^2 r(q^2/k^2), \quad
R_k^{AA}{}_{\mu\nu}(q)=Z_{\text{F}} {q^2 \, \left[\left(\delta_{\kappa\lambda}-\frac{q_\kappa q_\lambda}{q^2}\right)+(\delta_{\kappa\mu}-n_\kappa n_\mu)\frac{q_\mu q_\nu}{\alpha\, q^2}(\delta_{\nu\lambda}-n_\nu n_\lambda)+\frac{n_\kappa n_\lambda}{\beta\, q^2}\right]}r(q^2/k^2),
\end{equation}
{
The photon propagator in Landau gauge then takes the form
\begin{equation}
{(\Gamma^{(2)}+R_k)^{-1})_{\kappa\lambda}^{AA}(q)}{\vert_{a =0}}=\frac{1}{Z_F q^2 \left(1+r(q^2/k^2) \right)}
\left(\delta_{\kappa \lambda} - \frac{1}{q^2- (n \cdot q)^2} \left( q_{\kappa} q_{\lambda}+q^2n_{\kappa}n_{\lambda}\right)+ \frac{n \cdot q}{q^2- (n \cdot q)^2}\left(n_{\kappa}q_{\lambda}+n_{\lambda}q_{\kappa} \right)\right),
\end{equation}}
\end{widetext}
where $r(y)$ denotes a regulator shape function, specifying the
details of the regularization scheme, see below. 
Evaluating the Wetterich equation and determining the flow of
$\Gamma_k$ from a UV cutoff $\Lambda$ down to $k=0$ corresponds to a
quantization of axion electrodynamics in the path-integral framework. If a cutoff $\Lambda\to \infty$
limit existed, axion electrodynamics would even be UV complete. In the
following, we determine the flow of $\Gamma_k$ truncated down to the
form of \Eqref{eq:trunc} as a nonperturbative approximation to the
full effective action.

\section{RG flow of axion electrodynamics}
\label{sec:IV}

We are interested in the RG flow of all scale-dependent parameters of
$\Gamma_k$ in \Eqref{eq:trunc}. Let us start with the flow of the wave
function renormalizations $Z_a$ and $Z_F$. Their flow can be extracted
by projecting the right-hand side of the Wetterich equation onto the
corresponding kinetic operators: 
\begin{eqnarray}
\label{dtZaPro}
\partial_t Z_a &=&
\frac{1}{\Omega}\left[\pdif{^2}{q^2}\int\dif^4{p}\fdif{^2\partial_t\Gamma_k}{a(p)\delta
    a(-q)}\right]_{a,A,q\to0}, \\
\label{dtZFPro}
\partial_t Z_F &=& \frac{1}{\Omega}\left[\frac{4}{3}\pdif{}{q^2}\int\dif^4{p}\,n_\kappa n_\lambda\fdif{^2\partial_t\Gamma_k}{A_\kappa(p)\delta A_\lambda(-q)}\right]_{a,A,q\to0},
\end{eqnarray}
where $\Omega$ denotes the spacetime volume. Similar projection
prescriptions also exist for the flow of the (unrenormalized or bare) axion mass and
axion-photon coupling,
\begin{eqnarray}
\label{MassPro}
\partial_t \bar{m}_k^2 &=&\frac{1}{\Omega} \left[\int\dif^4{p}\fdif{^2 {\partial_t}\Gamma_k}{a(p)\delta a
    (-q)}\right]_{a,A,q\to0},\\
\label{KopplPro}
\partial_t \bar{g}_k &=& \frac{1}{\Omega}{\frac{i}{24}\, \epsilon_{\alpha\,\,\,\beta}^{\,\, \kappa \, \, \, \lambda}} \frac{\partial}{\partial r_{\alpha}} \frac{\partial}{\partial q_{\beta}} \nonumber\\
&& \times \int d^4p \left[\fdif{^3 {\partial_t}\Gamma_k}{a(p)\delta
    A_\kappa(r)\delta A_\lambda(-q)}\right]\Bigg|_{a,A,q\to0}\!\!.
\end{eqnarray}
It is straightforward to verify that {-- within our truncation -- } the RG flow of the latter
quantities vanishes exactly,
\begin{equation}
\pat \bar{m}_k^2=0, \quad \pat \bar{g}_k=0.\label{eq:nonrenorm}
\end{equation}
The reason for this nonrenormalization property is obvious from the
structure of the axion-photon vertex. From the first term of
\Eqref{GammaAA}, we observe that the vertex with an external axion
field has a nontrivial momentum structure $\sim q\, p\, a(q-p)$. Together
with the anti-symmetric Lorentz structure this vertex can only
generate operators containing derivatives of the axion field $\sim
\partial a$. This also ensures that no non-derivative axion-fermion couplings can be generated by integrating out photon fluctuations, as is in accordance with the axion being the pseudo-Goldstone boson of a spontaneously broken symmetry. In other words, an axion mass term or a higher-order
self-interaction potential cannot be generated {directly} from the axion-photon
vertex. 
A simple proof of \Eqref{eq:nonrenorm} can be based on the observation
that both the axion mass term as well as the axion-photon coupling are
operators which are nonzero for constant axion fields,
$a=a_{\text{c}}$. Hence, a projection of the right-hand side of the
flow equation onto $a=a_{\text{c}}$ suffices in order to extract the
flow of these operators. However, for a constant axion, the axion-photon
coupling can be written as 
\begin{equation}
\label{NonFPot}
\begin{aligned}
\int\dif^4{x}\,a_c F_{\mu\nu}\tilde{F}_{\mu\nu} &= 2 a_c \int\dif^4{x}\,\partial_\mu\left(A_\mu \tilde{F}_{\mu\nu}\right),
\end{aligned}
\end{equation}
i.e., as a total derivative of an abelian Chern-Simons
current. Therefore, it can be eliminated from the action and thus
cannot contribute to the renormalization flow of the nonderivative
axion terms of the theory.\footnote{{A similar observation was
    made in \cite{Reuter:1996be} in order to inversely argue that the
    renormalization flow of the topological charge in a non-abelian
    gauge theory can be properly formulated if the topological charge
    $\theta$ is temporarily considered as a spacetime dependent
    field.}} 

The nonrenormalization property extends to a full axion potential
\begin{equation}
\pat V_k(a^2) =0,
\end{equation}
which holds as long as the potential is purely mass-like at some
initial scale $\Lambda$, $V_\Lambda(a^2)= \frac{1}{2} \bar{m}_\Lambda^2
a^2$. Of course, as soon as axion self-interactions are present at
some scale, all parameters of the potential get renormalized through
these self-interactions. In other words, the pure axion mass term and
the axion-photon coupling are partial fixed points of the RG flow of
axion electrodynamics in terms of bare quantities. 

\begin{figure}
\includegraphics[width=\linewidth]{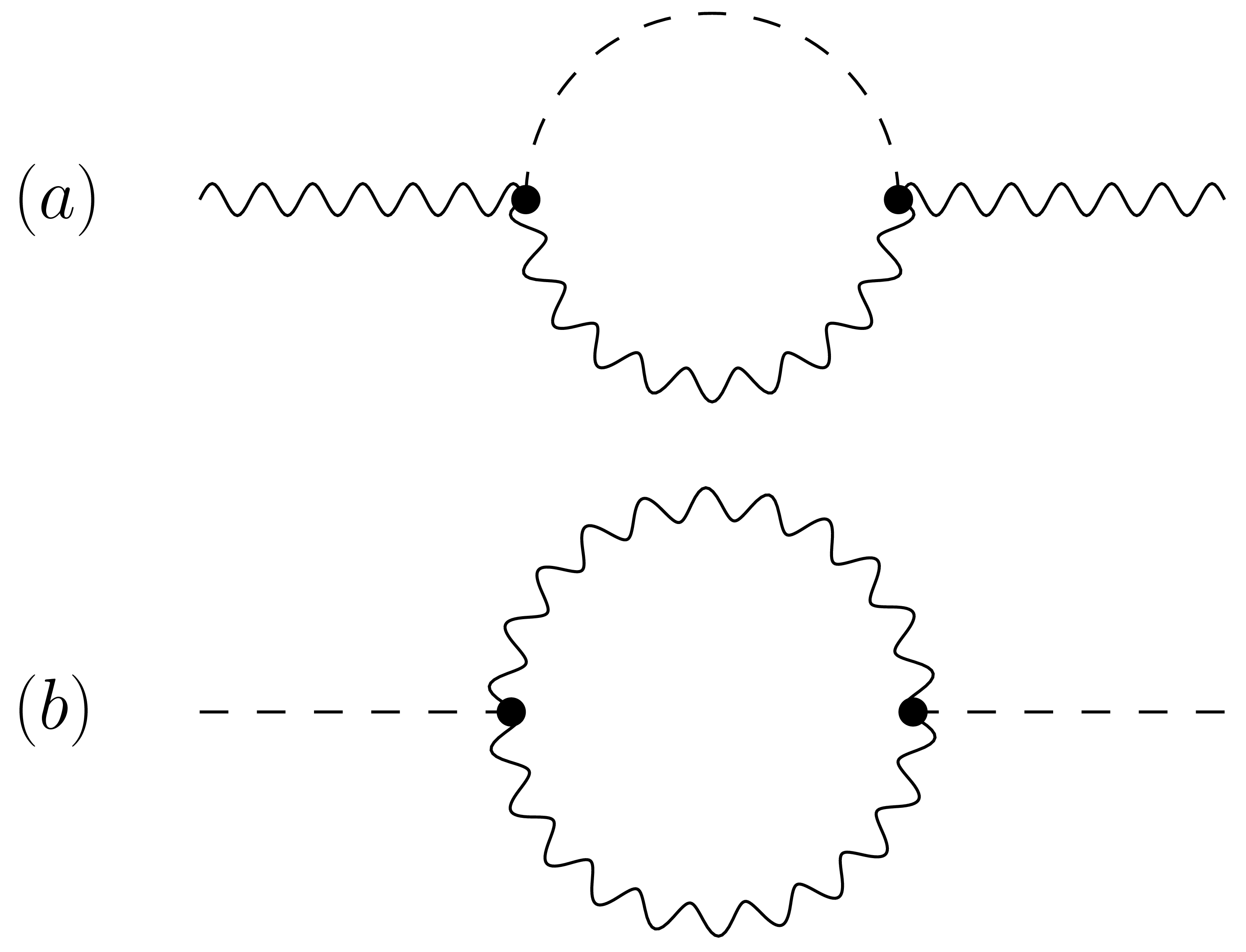}\\
\caption{\label{fig:diag} {Schematic representation of the
    diagrammatic contributions to $\eta_F$ (a) and $\eta_a$
    (b). Dashed lines denote axions, curly lines photons. As implied
    by the exact flow equation, all internal propagators and vertices
    are considered as fully dressed at the scale $k$. Appropriate
    insertions of $\partial_t R_k$ at one of the internal lines in
    each diagram and a corresponding sum over insertions is understood
    implicitly. The contribution to $\eta_F$ is thus to be understood
    as two distinct diagrams, associated with the contributions $\sim
    \eta_a$ and $\sim \eta_F$ in \Eqref{eq:etaF}.}}
\end{figure}

The only renormalization effects in our truncation therefore arise
from the running wave function renormalizations. The physical
observables can be expressed in terms of the (dimensionful)
renormalized couplings, given by
\begin{equation}
m_{\text{R}}^2 = \frac{\bar{m}_k^2}{Z_a}, \quad g_{\text{R}}^2= \frac{\bar{g}_k^2
  }{Z_F^2 Z_a}.\label{eq:rencoup}
\end{equation}
In order to investigate the structure of the RG flow and in particular
to search for fixed points at which the theory becomes scale-free, it is
convenient to introduce the renormalized dimensionless axion mass and
axion-photon coupling,
\begin{equation}
m^2 = \frac{\bar{m}_k^2}{k^2 Z_a}=\frac{m_{\text{R}}^2}{k^2}, \quad g^2= \frac{\bar{g}_k^2
  k^2}{Z_F^2 Z_a} = g_{\text{R}}^2 k^2.
\end{equation}
Defining the anomalous dimensions of the axion and the photon field
\begin{equation}
\eta_a=-\pat \ln Z_a, \quad \eta_F= - \pat \ln Z_F, 
\end{equation}
the $\beta$ functions of mass and coupling can be written as
\begin{eqnarray}
\pat g^2 &=& \beta_{g^2}= (2+2\eta_F+\eta_a) g^2, \label{eq:betag}\\
\pat m^2 &=& \beta_{m^2}=(\eta_a-2) m^2.\label{eq:betam}
\end{eqnarray}
From the projections \eqref{dtZaPro} and \eqref{dtZFPro} of the flow
onto the kinetic operators, the anomalous dimensions can be extracted:
\begin{eqnarray}
\eta_a &=& \frac{g^2}{6 (4\pi)^2} \left(2-\frac{\eta_F}{4} \right) , \label{eq:etaa}\\
\eta_F&=&  \frac{g^2}{6 (4\pi)^2} \left(
\frac{(2-\frac{\eta_a}{4})}{(1+m^2)^2} +
\frac{(2-\frac{\eta_F}{4})}{1+m^2} \right).\label{eq:etaF}
\end{eqnarray}
For these specific forms, we have used the linear regulator shape
function $r(y) = (\frac{1}{y} -1) \theta(1-y)$ \cite{Litim:2001up}. The
corresponding results for arbitrary shape functions are given in
Appendix \ref{app:A}. These are the central results of the present
paper, diagrammatically represented in
Fig. \ref{fig:diag}.

As axion electrodynamics belongs to the perturbatively
non-renormalizable theories according to naive power-counting, the
above-given $\beta$ or $\eta$ functions do not exhibit the same degree
of universality as $\beta$ functions of marginal couplings in
renormalizable theories. This implies that typically even the
leading-order (i.e., ``one-loop'') $\beta$ function coefficients are
scheme dependent. In the present case, we can change the numerical
value of the prefactors in Eqs.~\eqref{eq:etaa} and \eqref{eq:etaF} by
varying the regulator shape function $r(y)$. Nevertheless, the sign of
the prefactors cannot be changed for admissible regulators as is
visible from the explicit representations given in Appendix
\ref{app:A}. Furthermore, even for theories exhibiting this type of
non-unversality in their $\beta$ functions, the existence of fixed
points and the critical exponents determining the universality class
of the fixed point are universal.

Let us first discuss the resulting $\beta$ functions in various simple
limits. At weak coupling, the anomalous dimensions are small,
$\eta\sim g^2$. This implies that the anomalous dimensions on the
right-hand sides of Eqs.~\eqref{eq:etaa} and \eqref{eq:etaF},
signaling a typical RG ``improvement'', can be neglected. 
Further assuming a heavy-axion limit, $m^2\gg 1$, corresponding to a
renormalized axion mass being larger than a given scale under
consideration, $m_{\text{R}}^2\gg k^2$, the axions decouple, yielding
\begin{equation}
\eta_F\to 0, \quad \eta_a\to \frac{g^2}{3 (4\pi)^2}, \quad
\text{for}\,\, m\gg 1, g\ll 1.
\end{equation}
The axion anomalous dimension remains finite due to the fact that the
massless photons never strictly decouple from the flow. In the present
weak-coupling heavy-axion limit, the remaining flow of the coupling
reduces to
\begin{equation}
\pat g^2 = \left( 2+ \frac{g^2}{3 (4\pi)^2} \right) g^2, \quad
\text{for}\,\, m\gg 1, g\ll 1,\label{eq:betawcha}
\end{equation}
which can be straightforwardly integrated from a high UV scale
$\Lambda$ to some low scale $k$. In terms of the dimensionful
renormalized coupling, the solution reads
\begin{equation}
g_{\text{R}}^2(k) =\frac{g_{\text{R}}^2(\Lambda)}{1+ \frac{1}{6 (4\pi)^2} (\Lambda^2-k^2) g_{\text{R}}^2(\Lambda)}, \quad
\text{for}\,\, m\gg 1, g\ll 1,\label{eq:wcha}
\end{equation}
where $g_{\text{R}}^2(\Lambda)$ denotes the initial coupling value at
the UV cutoff. This value has to satisfy the weak coupling condition
$g_{\text{R}}^2(\Lambda) \Lambda^2 \ll 1$ . In the present
weak-coupling heavy-axion limit, we conclude that the photon-axion
coupling undergoes a finite renormalization even in the deep infrared
(IR) limit, despite the fact that photonic fluctuations never
{strictly} decouple in the deep IR.{ The photons still {\em
    effectively} decouple, as their low-momentum contributions to the
  flow vanish according to the powerlaw $\sim k^2
  g_{\text{R}}^2(\Lambda)$ for $k\to0$}. The same conclusion holds for the axion
mass. Inserting the coupling solution \eqref{eq:wcha} into the flow
for the mass \eqref{eq:betam}, the solution for the renormalized mass
reads
\begin{equation}
m_{\text{R}}^2(k)= \frac{m_{\text{R}}^2(\Lambda)}{{1+
    \frac{1}{6(4\pi)^2} (\Lambda^2-k^2) g_{\text{R}}^2(\Lambda)}},  \quad
\text{for}\,\, m\gg 1, g\ll 1,\label{eq:mwcha}
\end{equation}
Axion electrodynamics therefore exhibits a remarkable IR
stability. Quantitatively, both the coupling as well as the axion mass
run to slightly smaller values towards the infrared in the present
limit.

Let us now turn to the massless axion limit. As is obvious from
\Eqref{eq:betam}, the massless theory is an RG fixed point,
$m_\ast=0$. This follows from the non-renormalization of the axion
potential, which also implies that the axion will presumably not
exhibit the fine-tuning problem typically associated with massive
scalars. As long as $\eta_a<2$, this fixed-point is IR attractive.
  For weak coupling, the anomalous dimensions reduce to
\begin{equation}
\eta_F\to\frac{2g^2}{3 (4\pi)^2} , \quad \eta_a\to \frac{g^2}{3 (4\pi)^2}, \quad
\text{for}\,\, m=0, g\ll 1,
\end{equation}
yielding the coupling flow 
\begin{equation}
\pat g^2 = \left( 2+ 5\frac{g^2}{3 (4\pi)^2} \right) g^2, \quad
\text{for}\,\, m=0, g\ll 1.
\end{equation}
Integrating the flow analogously to \Eqref{eq:wcha} leads us to
\begin{equation}
g_{\text{R}}^2(k) =\frac{g_{\text{R}}^2(\Lambda)}{1+ \frac{5}{6 (4\pi)^2} (\Lambda^2-k^2) g_{\text{R}}^2(\Lambda)}, \quad
\text{for}\,\, m=0, g\ll 1.
\end{equation}
The conclusion is similar to the heavy-axion case: even though there
is no decoupling of any massive modes, axion electrodynamics shows a
remarkable IR stability. The fluctuations of the massless degrees of
freedom induce only a finite renormalization of the axion-photon
coupling yielding smaller couplings towards the IR. 

In the intermediate region for finite but not too heavy masses,
$m_\ast=0$ remains an IR attractive fixed point for weak coupling, but
this massless point is so weakly attractive that decoupling of the
axions typically sets in first and the flow ends up in the heavy-axion limit.

Let us now turn to arbitrary values of the coupling. For this {purpose}, we
solve Eqs. \eqref{eq:etaa} and \eqref{eq:etaF} for $\eta_a,\eta_F$ and
insert the result into Eqs. \eqref{eq:betag} and \eqref{eq:betam},
yielding the obviously nonperturbative flow equations
\begin{widetext}
\begin{eqnarray}
\label{ReDg}
\partial_t g^2 &=&\beta_{g^2}=2 g^2\frac{13g^4-384\pi^2g^2(21+17m^2+4m^4)-147456\pi^4(1+m^2)^2}{g^4-384\pi^2g^2(1+m^2)-147456\pi^4(1+m^2)^2},\\
\label{ReDm}
\partial_t m^2 &=&\beta_{m^2}=6m^2\frac{-g^4+128\pi^2g^2(4m^4+7m^2+3)-49152\pi^4(1+m^2)^2}{-g^4+384\pi^2g^2(1+m^2)+147456\pi^4(1+m^2)^2}.
\end{eqnarray}
\end{widetext}
In the heavy-axion limit the decoupling of the axion fluctuations
still simplifies the system considerably, since the suppression of
$\eta_F(m\to\infty)\to 0$ still persists, leading again to
\begin{equation}
\pat g^2 = \left( 2+ \frac{g^2}{3 (4\pi)^2} \right) g^2, \quad
\text{for}\,\, m\gg 1,\label{eq:betaha}
\end{equation}
as in \Eqref{eq:betawcha}. Also the corresponding mass flow is
equivalent to that of the heavy-axion weak-coupling limit. The
resulting flow exhibits no signature of a fixed point apart from the
Gau\ss ian one at $g=0$. On the contrary, the $\beta$ function for the
running coupling is {somewhat} similar to that of one-loop QED. The
integrated flow will therefore give rise to a Landau pole of the
running coupling at high scales. Fixing the renormalized dimensionful
coupling to some value at the scale $k=0$, i.e.,
$g_{\text{R}}^2(k=0)=g_{\text{R}0}^2$, the scale of the Landau pole
$\mu_{\text{L}}$, where $g_{\text{R}}^2(\mu_{\text{L}})\to \infty$,
can directly be read off from \Eqref{eq:wcha},
\begin{equation}
\mu_{\text{L}}\simeq \frac{\sqrt{6}(4\pi)}{g_{\text{R}0}} \simeq
\frac{31}{g_{\text{R}0}},
\label{eq:muL1}
\end{equation}
which is an order of magnitude larger than the inverse coupling in the
deep IR. For this estimate to hold, the dimensionless mass has to be
large during the whole flow. With regard to \Eqref{eq:mwcha}, this is
true as long as $1\ll m^2 = m_{\text{R}}^2/k^2$ is satisfied for all
$k$. This is always true if $m_{\text{R}}^2(k=0)/\mu_{\text{L}}^2\gg
1$ (this criterion could even be relaxed a bit). Going towards larger
scales, $m_{\text{R}}(k)$ increases with $k$ and diverges even {at
$k=\mu_{\text{L}}$.}

\begin{figure}[!here]
\includegraphics[width=\linewidth]{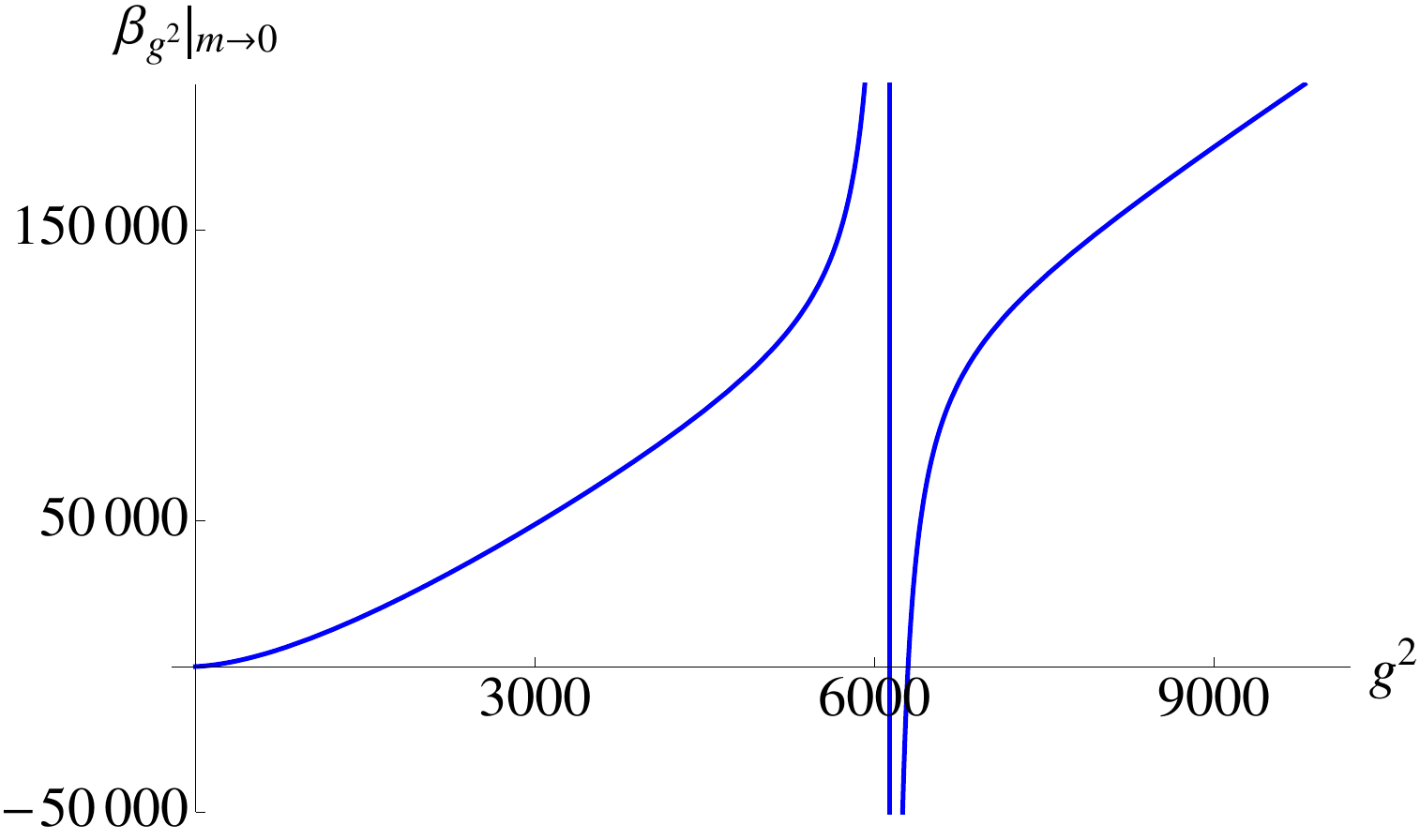}
\caption{\label{fig:betagplot}The $\beta$ function $\beta_{g^2}$ in the massless limit clearly exhibits an IR-attractive GFP, a singularity at $g=g_{\rm sing}$, and an asymptotic behavior $\sim 26 g^2$.}
\end{figure}

A different structure becomes visible in the massless limit for
arbitrary coupling. Whereas $m_\ast=0$ still is a fixed point of the
RG, the nonperturbative flow of the coupling becomes
\begin{equation}
\label{Regls}
\pat g^2= 2g^2\frac{13g^4-8064\pi^2
  g^2-147456\pi^4}{g^4-384\pi^2{g^2}-147456\pi^4},\quad
\text{for}\,\, m=0.
\end{equation}
Starting from small coupling values, the $\beta_{g^2}$ function runs
into a singularity ($\beta_{g^2} \to \infty$) for 
\begin{equation}
g_{\text{sing}}^2= 64 \pi^2 (3+3\sqrt{5}), \quad \text{for}\,\, m=0,
\end{equation}
where the denominator changes sign. For even larger couplings, the
$\beta_{g^2}$ function returns from $-\infty$ and has a zero at
\begin{equation}
g_\ast^2=64 \pi^2\left( \frac{3(21+\sqrt{493})}{13} \right), \quad \text{for}\,\, m=0.
\end{equation}
For $g^2\to \infty$, the $\beta_{g^2}$ function approaches the simple
form
\begin{equation}
\pat g^2 \simeq 26 g^2,\quad \text{for}\,\, g^2\to \infty, m=0.
\end{equation}
The massless $\beta_{g^2}$ function is plotted in
Fig. \ref{fig:betagplot}. \\

Fixing the renormalized dimensionful coupling in the deep IR as above,
$g_{\text{R}}^2(0)=g_{\text{R}0}$, we can integrate the flow until the running
coupling hits the singularity of the $\beta_{g^2}$ function. The
corresponding scale $\mu_{\text{L}}'$ can be viewed as an
``RG-improved'' Landau pole, giving a nonperturbative estimate of the
scale of maximum UV extension of quantum axion electrodynamics. From a
numerical integration, we obtain
\begin{equation}
\mu_{\text{L}}'\simeq \frac{14.844}{g_{\text{R}0}}, \quad
\text{for}\,\, m=0.
\label{eq:muL2}
\end{equation}
This is of the same order of magnitude as the scale $\mu_{\text{L}}$
found above for the heavy-mass limit. {As a rough estimate of the regularization-scheme dependence of this result, let us note that in the case of an exponential shape function $r(y)=({\rm exp}(y)-1)^{-1}$ we obtain $\mu_{\rm L\, exp}' \simeq \frac{10}{g_{\rm R 0}}$.} 

For finite values of the axion mass, the flow of the coupling
generically exhibits the same features as for the massless case. The
position of the singularity in $\beta_{g^2}$ is shifted towards larger
values of the coupling for increasing mass. It is straightforward to
verify that the $\beta$ functions \eqref{ReDg}, \eqref{ReDm} do not
support a nontrivial fixed point for physically admissible positive
values of $g^2$ and $m^2$. At the same time, the mass flow exhibits
the same singularity in $\beta_{m^2}$ as the denominators in
Eqs. \eqref{ReDg} and \eqref{ReDm} are identical.

We conclude that generic flows in the physically admissible parameter
space cannot be extended beyond a scale of maximum UV
extension. {Phenomenological implications of the existence of
  such a maximum UV scale  will be discussed in the next
  section.}  Whether or not substantial extensions of our truncation
are able to modify this conclusion is hard to predict. For instance,
an inclusion of axion self-interactions would add another sector to
the theory which generically suffers from a triviality problem and is
thus not expected to change our conclusions. {Qualitative
  modifications of our} {$\beta$ functions could
  {potentially} arise from momentum-dependent axion
  self-interactions.  These are induced by photon fluctuations even
  within our truncation, and can couple into the flow of $\eta_a$,
  thus {also contributing to} the $\beta$ functions for the axion
  mass and axion-photon coupling, as well as further
  {momentum-dependent} axion self interactions.} 

 For the remainder of this section, let us concentrate on an oddity of
 the present flow. There is in fact one exceptional RG trajectory
 which can be extended to all scales. This trajectory requires the
 singularity induced by the denominator of the $\beta$ function to be
 canceled by a zero of the numerator. Remarkably, there exists a mass
 and coupling value in the physically admissible region, where the
 singularities in both the mass and the coupling flow are
 canceled. This exceptional point in theory space is given by
\begin{equation}
m_{\text{exc}}^2 =\frac{1}{2} (\sqrt{5}-1), \quad
\frac{g_{\text{exc}}^2}{6 (4\pi)^2} = 2(3+\sqrt{5}). \label{eq:excp}
\end{equation}
The RG trajectory which passes through this point, say at a scale
$\Lambda_{\text{exc}}$, has a standard IR behavior exhibiting the
typical decrease towards smaller renormalized mass and
couplings. Towards the UV, the $\beta$ functions do not exhibit a
fixed point but approach the simple form
\begin{equation}
\pat g^2 \simeq 26 g^2, \quad \pat m^2 \simeq 6 m^2,
\end{equation}
which resembles a pure dimensional scaling with large anomalous
dimensions.
In fact, this behavior corresponds to $\eta_a \to 8$, and
$\eta_F \to 8$. It implies that both dimensionless and dimensionful
renormalized couplings increase strongly towards the UV without
hitting a Landau pole singularity. Instead the growth of the couplings
remains controlled on all scales and approaches infinity at infinite
UV cutoff scale. 

{As expected, the scaling dimensions are non-universal, and in fact show a considerable regulator dependence: Employing an exponential shape function yields $\pat g^2 \simeq \frac{31}{2} g^2$ and $\pat m^2 \simeq \frac{5}{2}m^2$, corresponding to $\eta_a = 4.5 = \eta_F$.}

An important consequence of this exceptional flow is that the physical
IR values are completely fixed in terms of the scale
$\Lambda_{\text{exc}}$. Numerically integrating the flow towards
$k\to0$ yields { for the linear regulator}
\begin{eqnarray}
g_{\text{R}}(k=0)&\equiv& g_{\text{R}0}\simeq
\frac{20.36}{\Lambda_{\text{exc}}},\label{eq:gexc}\\
 \quad m_{\text{R}}(k=0)&\equiv& m_{\text{R}0}\simeq
0.4657\Lambda_{\text{exc}}, \nonumber\\
\end{eqnarray}
implying the dimensionless combination
$g_{\text{R}0}m_{\text{R}0}\simeq 9.484$. {Of course, our observation
  of this exceptional trajectory requires a critical discussion: The
  fact that the anomalous dimensions become comparatively large may be
  interpreted as a signature that explicit momentum-dependencies of
  the propatators and vertices become important at larger coupling. If
  so, the exceptional trajectory might just be an artefact of our
  truncation which assumes tree-level-type propagators and
  vertices. On the other hand, we cannot exclude the possibility that
  this exceptional trajectory is a simple projection of a legitimate
  UV-extendable trajectory in the full theory space. If so, the
  observed exceptionality could reflect the restrictions on the
  physical parameters induced by the true UV behavior (potentially
  controlled by a UV fixed point). Our truncated RG flow could then be
  quantitatively reliable below $\Lambda_{\text{exc}}$.} {The
  strong regulator dependence observed above should be read as a hint
  that if such an exceptional trajectory exists within the full theory
  space, it might still change considerably when operators beyond our
  present truncation are taken into account.} {In particular, the
  large anomalous dimensions seem to call for the inclusion of
  higher-derivative operators.}

{In the rather speculative case, that the trajectory exists with the
  same qualitative features in the full theory, we would have
  discovered a (to our knowledge) first example of a UV complete
  theory with a high energy behavior that is controlled by
  {neither} a fixed point nor a limit cycle, see,
    e.g., \cite{Litim:2012vz}. Though this theory would not fall into
  the class of asymptotically safe systems due to the lack of a UV
  fixed point, it would be asymptotically controllable. The number of
  physical parameters of such systems would then correspond to the
  dimensionality of the exceptional manifold, i.e., the analog of
  \Eqref{eq:excp} including all possible further couplings. We
  emphasize that in this speculative case also further properties
  required for a legitimate field theory such as unitarity would have
  to be critically examined.}  { In particular, a large positive
  anomalous dimension, as observed here, implies a strongly
  UV-divergent propagator $\sim (p^{2})^{1- \eta/2}$, which might
  result in cross-sections increasing as a large power of the
  momentum.  Whether such behavior can be reconciled with requirements
  such as perturbative unitarity within standard quantum field theory
  remains to be investigated.}

\section{Phenomenological implications}
\label{sec:implications}

\subsection{QCD axion}
Axion electrodynamics occurs naturally as a low-energy effective
theory in the context of the Peccei-Quinn solution of the strong CP
problem. The QCD axion develops a generic two-photon coupling, induced
by its mixing with the mesonic $\pi^0$, $\eta$ {and $\eta'$} degrees of
freedom. Effective axion electrodynamics therefore arises during the
chiral phase transition of QCD at around a typical QCD scale which we
choose to be $\Lambda_{\text{QCD}}\simeq 1$ GeV. At that scale, the
relation between axion coupling and mass is essentially fixed by the
corresponding pion scales, 
\begin{equation}
g_{\text{R}}(\Lambda_{\text{QCD}}) = C\, \frac{
  m_{\text{R}}(\Lambda_{\text{QCD}})}{m_\pi f_\pi}, 
\label{eq:QCDaxion}
\end{equation}
where $m_\pi$ and $f_\pi$ denote the pion mass and decay constant
respectively, and $C$ is a numerical dimensionless constant that
depends of the microscopic axion model (typically defined in the
context of a grand unified scenario). For instance, for the KSVZ
axion \cite{Kim:1979if}, the coupling mass relation yields
$g_{\text{R}}(\Lambda_{\text{QCD}}) \simeq \frac{0.4}{\text{GeV}^2}
m_{\text{R}}(\Lambda_{\text{QCD}})$. Other axion models typically lie
within an order of magnitude near this relation
\cite{Dine:1981rt,Cheng:1995fd}. 

\begin{figure}[!here]
\includegraphics[width=\linewidth]{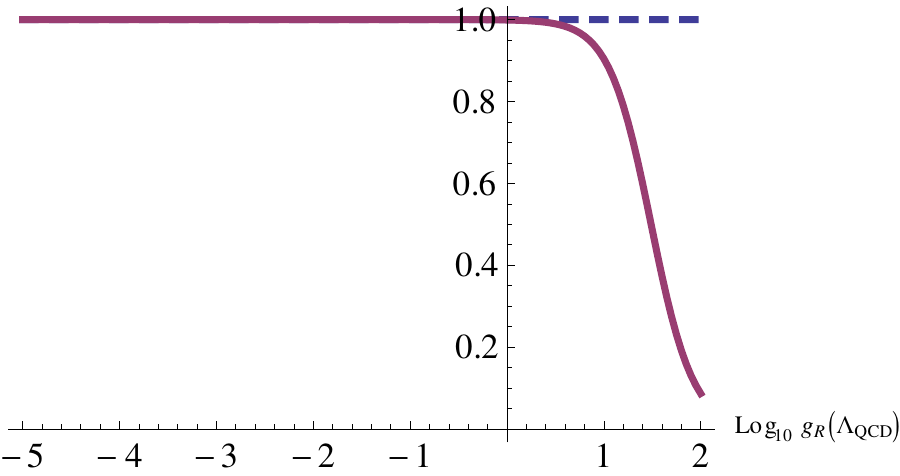}\\
\caption{\label{fig:ratios} Plot of the renormalized coupling ratios
  $c_{\text{R}}$ of \eqref{eq:cR} (blue/dashed line) and $p_{\text{R}}$ of
  \eqref{eq:pR} (purple/solid line) as a function of the logarithm of the
  initial axion-photon coupling $g_{\text{R}}(\Lambda_{\text{QCD}})$
  in units of GeV. For small couplings, the renormalization of both
  coupling and mass remain unobservably small. Only for larger
  couplings, the renormalization of both towards smaller values
  becomes visible (purple/solid line), while the proportionality between
  mass and coupling remains essentially unaffected ({blue}/dashed line).}
\end{figure}

Axion searches in experiments or astrophysical/cosmological
observations do actually not test the parameters occurring in
\Eqref{eq:QCDaxion} directly, as they do not operate near the QCD
scale, but typically at much lower scales $k_{\text{obs}}$. They range from $\sim$keV
momentum scales of stellar evolution theory of horizontal-branch
stars, via $\sim$eV scales for solar energy loss or direct helioscope
observations, to $\sim\!\mu$eV scales in light-shining-through walls
experiments. In other words, a proper comparison of such observational
results with \Eqref{eq:QCDaxion} requires to take the finite
renormalization of axion electrodynamics between
$\Lambda_{\text{QCD}}$ and $k_{\text{obs}}$ into account. For
simplicity, we estimate the maximum renormalization effect by
choosing $k_{\text{obs}}=0$. In order to illustrate our findings, we
plot in Fig.~\ref{fig:ratios} the renormalized ratio
\begin{equation}
c_{\text{R}}=
\frac{g_{\text{R}}(k_{\text{obs}})/m_{\text{R}}(k_{\text{obs}})}{g_{\text{R}}(\Lambda_{\text{QCD}})/m_{\text{R}}(\Lambda_{\text{QCD}})}
\label{eq:cR}
\end{equation}
inspired by \Eqref{eq:QCDaxion} as well as the ratio of the product
\begin{equation}
p_{\text{R}}=
\frac{g_{\text{R}}(k_{\text{obs}})m_{\text{R}}(k_{\text{obs}})}{g_{\text{R}}(\Lambda_{\text{QCD}})m_{\text{R}}(\Lambda_{\text{QCD}})}
\label{eq:pR}
\end{equation}
as {dark-blue (dashed)}  and purple {(solid)} lines, respectively. The ratios are plotted
 as a function of the  logarithm of the initial coupling
$g_{\text{R}}(\Lambda_{\text{QCD}})$ in units of GeV.  For small
couplings, the renormalization of both coupling and mass remain
unobservably small implying that both ratios
$c_{\text{R}},p_{\text{R}}\to 1$. Only for larger couplings, the renormalization of
both towards smaller values becomes visible, implying a significant
decrease of $p_{\text{R}}$. Most importantly, the
proportionality between mass and coupling remains essentially
unaffected. Numerically, the ratio $c_{\text{R}}$ changes only on the
$10^{-5}$ level.

We conclude that exclusion bounds derived from the non-observation of
axion effects are not modified by renormalization effects of axion
electrodynamics along the lines of constant
$g_{\text{R}}/m_{\text{R}}$. As the physically relevant parameter
space for the QCD axion lies below $g_{\text{R}}\lesssim 10^{-9}
(\text{GeV})^{-1}$, we conclude from Fig.~\ref{fig:ratios} that the
renormalization effects within the effective theory of axion
electrodynamics are completely irrelevant.

{The existence of a scale of maximum UV
  extension $\mu_{\text{L}}$ also seems to be of no relevance for the
  QCD axion: for instance, for an axion coupling in the range of $g
  \sim \mathcal{O}(10^{-16}) \rm GeV$ where the axion could also
  provide a substantial part of the dark matter of the universe, the
  scale $\mu_{\text{L}}$ would lie well beyond a typical GUT scale. In
  turn, this demonstrates that axion electrodynamics is a consistent
  effective (quantum) field theory at low energies within the QCD
  axion scenario.}

For completeness, let us mention that the QCD axion would be on the
exceptional trajectory found above, if
$g_{\text{R}}(\Lambda_{\text{QCD}})\simeq 1.9 \,\text{GeV}^{-1}$ and 
$m_{\text{R}}(\Lambda_{\text{QCD}})\simeq 5.0\, \text{GeV}$. This is, of
course, beyond the parameter regime conventionally considered for the
QCD axion. 

\subsection{General axion-like particles (ALPs)}

Apart from the QCD axion, massive scalar bosons may arise as
pseudo-Nambu-Golstone bosons in a variety of contexts
\cite{Wilczek:1982rv,Chikashige:1980ui}. In case of a pseudo-scalar
boson, the emergence of a coupling to photons of 
axion-electrodynamics-type is a natural consequence. Since the
coupling-mass relation in this general case is not necessarily fixed
as in \Eqref{eq:QCDaxion}, a more general parameter space for such
axion-like-particles can be investigated.

A priori, the mass-coupling space seems rather unrestricted. Depending
on the mass-generation mechanism, the ALP mass can vary strongly from
very light masses arising from anomalous breakings as in the QCD-axion
case, to explicit masses of the order of the high scale
$\Lambda_{\text{UV}}$ (such as the Fermi, GUT, or Planck
scale). Depending on the value of the coupling, the low-energy mass
may be very different from the high-energy mass due to
renormalizations of the type \eqref{eq:mwcha}, but, in general, the
renormalized low-energy mass still remains rather unconstrained.

This can be different for the possible values of the effective low-energy
ALP-photon coupling. If the effective low-energy theory still exhibits
a scale of maximum UV extension $\mu_{\text{L}}$ as in Eqs.~\eqref{eq:muL1} or
\eqref{eq:muL2}, then that scale must necessarily be higher than the
microscopic high energy scale $\Lambda_{\text{UV}}$. Otherwise, the effective low-energy
theory could not possibly arise from that unknown microscopic
theory. As a consequence, the low-energy coupling is bounded from
above due to the renormalization flow in the effective theory,
\begin{equation}
\text{max}\, g_{\text{R}}(k\to 0)\equiv g_{\text{R}0,\text{max}} =
\frac{N_{\text{R}}}{\mu_{\text{L}}}\leq 
\frac{N_{\text{R}}}{\Lambda_{\text{UV}}},
\label{eq:impbound}
\end{equation}
where $N_{\text{R}}$  is a number that depends on the ALP mass as well
as on further degrees of freedom in the effective low-energy
theory. In the case that this effective theory is well-approximated by
axion electrodynamics, we found that $N_{\text{R}}=\mathcal{O}(10)$
($N_{\text{R}}\simeq 31$ for large masses and $N_{\text{R}}\simeq
14.8$ for the massless case). 

Our conclusion, that $g_{\text{R}0}$ should be suppressed by the high
scale $\Lambda_{\text{UV}}$ looks rather trivial, as it seems to
follow standard power-counting arguments for a higher-dimensional
operator. However, we stress that the statement is actually stronger:
even for unusually enhanced couplings at the high scale (invalidating
naive power-counting), axion electrodynamics as an effective theory
ensures that the low-energy coupling is supressed by renormalization
effects and obeys the bound \eqref{eq:impbound}.

We expect the precise number $N_{\text{R}}$ to be modified by further
degrees of freedom, such as the standard-model fermions contributing
to the flow at higher scales. As long as
they leave the anomalous dimension $\eta_a$ in \Eqref{eq:etaa}
positive, we expect a mere quantitative influence on
$N_{\text{R}}$. Of course, these conclusions no longer hold, if the
additional degrees of freedom render the effective theory
asymptotically free or safe, or if the system sits on the exceptional
trajectory as discussed above.

\section{Conclusions}
\label{sec:conclusions}

We have analyzed the renormalization flow of axion electrodynamics
considered as an effective quantum field theory. From a field-theory
viewpoint, we have revealed several interesting properties:
non-renormalization properties protect the flow of the axion mass and
axion-photon coupling, such that {within our truncation} the flow of the corresponding
renormalized quantities is solely determined by the axion
and photon anomalous dimensions. These non-renormalization properties
are in line with but go beyond the fact that the axion can be
understood as a pseudo-Nambu-Goldstone boson of a broken Peccei-Quinn
symmetry. 

Towards the infrared, the flow remains well controlled even in the
presence of massless degrees of freedom. Even though massless photons
strictly speaking never decouple, their contribution to the RG flow
effectively decouples leading to finite and predictable IR
observables. By contrast, the UV of axion electrodynamics exhibits a triviality
problem somewhat similar to $\phi^4$ theory: insisting on sending the
cutoff $\Lambda\to\infty$ is only possible for the free theory $g\to
0$. From a more physical viewpoint, fixing the coupling to a finite
value at a finite scale implies that axion electrodynamics can be
treated as a quantum field theory only up to a scale of maximum UV
extension $\mu_{\text{L}}$. The value of this scale is, of course, not
universal. For our regularization scheme, this scale is about an order
of magnitude larger than the inverse IR coupling and depends
weakly on the axion mass. 

A behavior different from this generic case is only found on an
exceptional RG trajectory which remains free of singularities on all {finite}
scales. From within our truncated RG flow, it is difficult to decide
whether this trajectory is a mere artefact of the truncated theory
space or a remnant of a valid trajectory of an interacting
UV-controlled theory. In the latter case, it could be the first
example of a predictive and consistent theory without being associated
with {an obvious} UV fixed point. At the present level of approximation, however,
we consider this exceptional trajectory as an oddity, the status,
physical relevance and consistency of which still has to be carefully
examined. 

For the QCD axion, our findings demonstrate that the typical
proportionality between axion-photon coupling and axion mass is not
relevantly renormalized by low-energy axion-photon fluctuations. In
fact, possible renormalizations of mass and coupling largely cancel
out of the proportionality relation even at stronger coupling. As a
consequence, phenomenological bounds on these parameters remain
essentially unaffected by the RG flow. By contrast, the absolute
values of mass or coupling can undergo a sizeable renormalization,
however, the required coupling strength is not part of the natural
QCD axion regime. 

For more general axion-like particles the existence of a generic
maximum scale of UV extension induces an upper bound on possible
values of the axion-photon coupling. Standard lines of perturbative
reasoning \cite{Georgi:1986df} suggest that the renormalized coupling
should be of order $g_{\text{R}}\sim 1/\Lambda_{\text{UV}}$, where
$\Lambda_{\text{UV}}$ denotes the microscopic scale where the axion
sector is coupled to the standard model particles. Our RG study now
demonstrates that couplings of that size are not only natural, but are
in fact bounded by $g_{\text{R}}\lesssim
\mathcal{O}(10)/\Lambda_{\text{UV}}$ due to renormalization effects in
the axion-photon sector. In view of the rather unconstrained ALP
parameter space at large masses and comparatively large couplings
(see, e.g., the compilation in \cite{Hewett:2012ns}), our bound could
become of relevance for ALP searches at hadron colliders above
$m_{\text{R}}\gtrsim 1$GeV and couplings above $g_{\text{R}}\gtrsim
10^{-3}/(\text{GeV})^{-1}$.

\acknowledgments The authors thank J.~Jaeckel for helpful discussions and acknowledge support by
  the DFG under grants Gi~328/1-4. HG thanks the DFG for support
  through grant Gi~328/5-2 (Heisenberg program) and SFB-TR18. 

Research at Perimeter Institute is supported by the Government of Canada through Industry Canada
and by the Province of Ontario through the Ministry of Research and Innovation.

\appendix

\section{RG flows for general regulator shape functions}
\label{app:A}

{For generality, we present the flows of the wave function
  renormalizations for general regulator shape functions $r(y)$. For
  the flow of the axion wave function, we obtain}
{
\begin{equation}
\partial_t Z_a= - \frac{g^2}{2 \left( 2 \pi\right)^4}\int d^4p\,
\frac{-\eta_F r_k\left( \frac{p^2}{k^2}\right)+ \partial_t r_k\left(
  \frac{p^2}{k^2}\right)}{Z_F^2 p^2 \left(1+r_k\left[
    \frac{p^2}{k^2}\right] \right)^3}. 
\end{equation}
}
{The flow of the photon-wave function is given by}
{
\begin{widetext}
\begin{eqnarray}
&{}&\partial_t Z_F\\
&=& - \frac{g^2}{2}\! \int\!\! \frac{d^4p}{ \left(2 \pi \right)^4} \left(\! \frac{-Z_a \eta_a p^2 r_k\left( \frac{p^2}{k^2}\right) + Z_a p^2 \partial_t r_k\left( \frac{p^2}{k^2}\right)}{\left( m^2 +Z_a p^2 \left[ 1+r_k\left( \frac{p^2}{k^2}\right)\right]\right)^2  Z_F \left[1+r_k\left( \frac{p^2}{k^2} \right]\right)}
+\frac{-\eta_F r_k\left( \frac{p^2}{k^2}\right)+ \partial_t r_k\left( \frac{p^2}{k^2}\right)}{Z_F \left[1+r_k\left( \frac{p^2}{k^2}\right) \right]^2 \left(m^2 + Z_a p^2\left[ 1+r_k \left( \frac{p^2}{k^2}\right)\right] \right)}
\!\right)\! .\nonumber
\end{eqnarray}
\end{widetext}}

{Inserting the linear regulator shape
function $r(y) = (\frac{1}{y} -1) \theta(1-y)$, the momentum integrals
can be performed analytically. The results can be expressed in terms
of the anomalous dimensions and are given by Eqs.~\eqref{eq:etaa} and
\eqref{eq:etaF}. }

\section{Euclidean axion electrodynamics with imaginary coupling}
\label{app:B} 

In the main text, we have emphasized that the physically admissible
parameter space is constrained to positive masses and couplings
(squared), $g^2,m^2>0$. For $m^2<0$ and in absence of any further
axion potential, the Euclidean action is unbounded from below along
the direction of large axion field amplitude. Of course, this could be
cured by adding a stabilizing potential, but this route will not be
followed in this work.

A more interesting case is provided by the case of imaginary
axion-photon couplings, $g^2<0$. In this case, the Euclidean action
would violate Osterwalder-Schrader reflection positivity, such that a
corresponding Minkowskian theory can be expected to violate
unitarity. Still, the Euclidean theory could be regarded as a valid
field theory description of some suitable statistical system. In this
case, the action would be stable along the axion-amplitude direction,
but unstable towards the formation of large electromagnetic fields
with large $\mathbf{E\cdot B}$. This instability could be cured by
higher photon self-interaction such as 
\begin{equation}
\mathcal{L}_{F} = f_1 (F_{\mu\nu}F_{\mu\nu})^2 +
f_2(F_{\mu\nu}\widetilde{F}_{\mu\nu})^2,
\end{equation}
with positive constants $f_{1,2}$ carrying a mass dimension of
$-4$. Actions of this type are familiar from fluctuation-induced
nonlinear QED contributions to electrodynamics
\cite{Heisenberg:1935qt,Dittrich:2000zu}. Also within axion
electrodynamics, we expect these contributions to be generated by
mixed axion-photon fluctuations (some properties of such amplitudes
have, for instance, been studied in \cite{Grifols:1999ku}). In the
following, we simply assume that these terms are suitably generated
either within axion electrodynamics or provided by an exterior sector
coupling to photons. Then, also the parameter region where $g^2<0$ can
become physically admissible.

{
In that case, the $\beta$ function for $\hat{g}=- i g$ is given by 
\begin{widetext}
\be
\beta_{\hat{g}^2}=2 \hat{g}^2\frac{13\hat{g}^4+384\pi^2\hat{g}^2(21+17m^2+4m^4)-147456\pi^4(1+m^2)^2}{\hat{g}^4+384\pi^2\hat{g}^2(1+m^2)-147456\pi^4(1+m^2)^2},\\
\label{beta_img}
\ee
The mass $\beta$ function now takes the form
\be
\beta_{m^2}=6m^2\frac{-\hat{g}^4-128\pi^2\hat{g}^2(4m^4+7m^2+3)-49152\pi^4(1+m^2)^2}{-\hat{g}^4-384\pi^2\hat{g}^2(1+m^2)+147456\pi^4(1+m^2)^2}.
\ee
\end{widetext}
In this system, we find a UV attractive fixed point at $(\hat{g},
m)\approx (13.25,0)$. The critical exponents, defined as eigenvalues
of the stability matrix $\frac{\partial \beta_{g_i}}{\partial g_j}$
with $g_i= (\hat{g}, m)$, multiplied by an additional negative sign,
are $(\theta_1, \theta_2)= (2.16, 1.20)$. Thus this fixed point is UV
attractive in both directions.}

{Clearly the system also admits a Gau\ss{}ian fixed point with critical
exponents given by the canonical dimensions. Accordingly this fixed
point is IR attractive in the coupling.}

\begin{figure}[t]
\includegraphics[width=\linewidth]{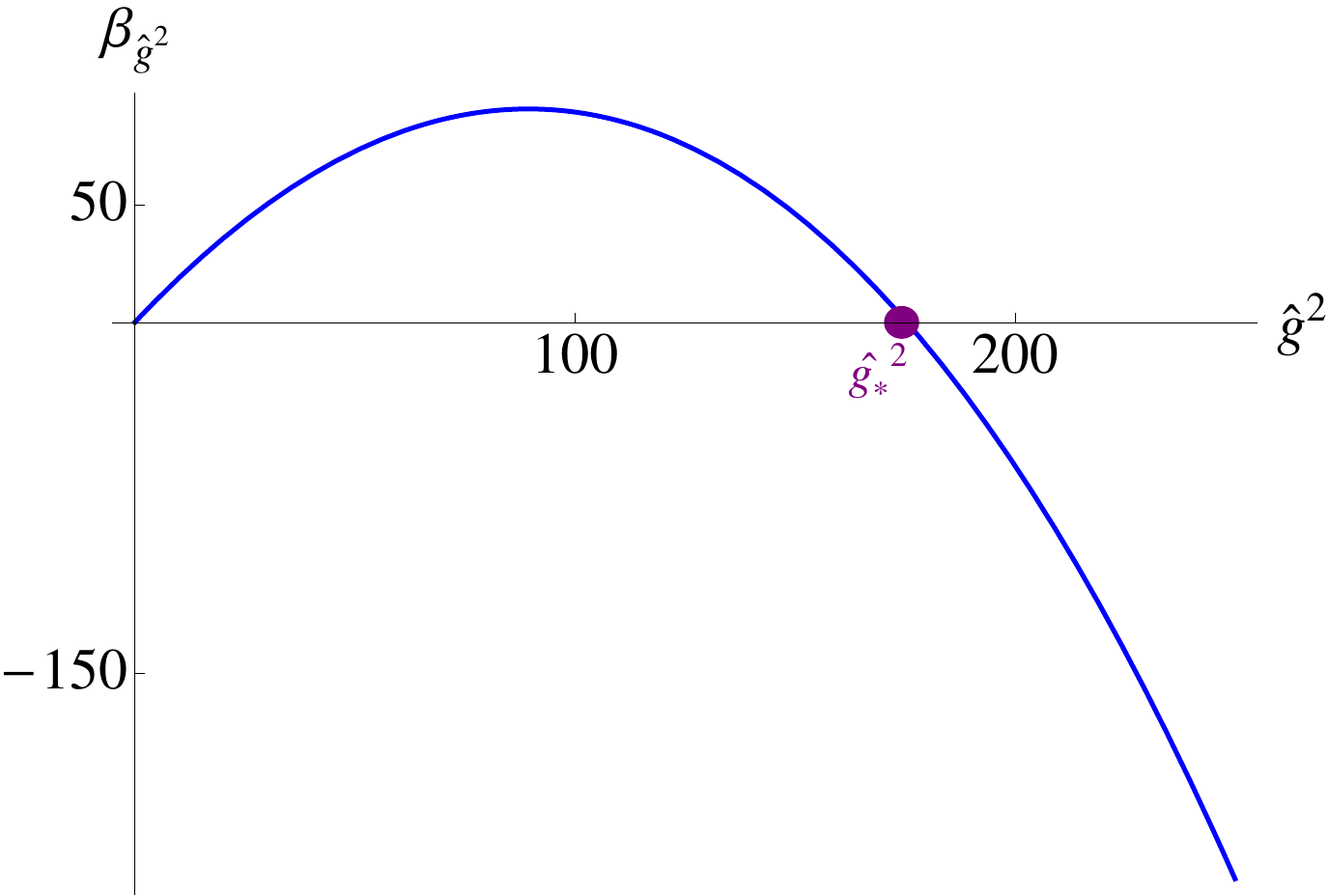}
\caption{\label{fig:beta_imgsq} The $\beta$ function for the coupling
  $\hat{g}$ in the limit $m \rightarrow 0$ clearly shows a UV
  attractive non-Gau\ss{}ian fixed point at $\hat{g}= \hat{g}_{\ast}=
  13.25$, as well as an infrared attractive Gau\ss{}ian fixed point.}
\end{figure} 

{ 
We conclude that, setting the axion mass to zero, the system admits
the construction of a complete RG trajectory, extending from
$\hat{g}=0$ at $k \rightarrow 0$ to $\hat{g}= \hat{g}_{\ast}$ for $k
\rightarrow \infty$, {see fig}.~\ref{fig:beta_imgsq}. Thus the axion-photon system with an imaginary
coupling {has the potential to provide} a simple example of an asymptotically safe quantum
field theory, albeit {without an immediate}  physical application.}


\end{document}